\documentclass[twocolumn,10pt]{IEEEtran}
\usepackage{algorithm,algorithmic,amsbsy,amsmath,amssymb,epsfig,bbm,mathrsfs,multirow,amsthm,subcaption,caption,graphicx,epstopdf,multicol,lipsum,float}
\usepackage{adjustbox}
\usepackage{hyperref}
\makeatletter
\g@addto@macro{\UrlBreaks}{\UrlOrds}
\makeatother

\hypersetup{
  colorlinks=true,
  linkcolor=blue,
  filecolor=blue,   
  urlcolor=black,citecolor=blue,
}
\usepackage{listings}
\usepackage{color}
\usepackage[english]{babel}
\usepackage{blindtext}

\definecolor{mygreen}{rgb}{0,0.6,0}
\definecolor{mygray}{rgb}{0.5,0.5,0.5}
\definecolor{mymauve}{rgb}{0.58,0,0.82}

\begin{document}

\title{Performance Analysis and Application of Mobile Blockchain}

\author{\IEEEauthorblockN
{Kongrath Suankaewmanee$^1$, Dinh Thai Hoang$^1$, Dusit Niyato$^1$, Suttinee Sawadsitang$^1$, Ping Wang$^1$, and Zhu Han$^2$\\
$^1$ School of Computer Science and Engineering, Nanyang Technological University, Singapore	\\
$^2$ Department of Electrical and Computer Engineering, University of Houston, USA 	\vspace{-5mm}}}

\maketitle

\begin{abstract}

Mobile security has become more and more important due to the boom of mobile commerce (m-commerce). However, the development of m-commerce is facing many challenges regarding data security problems. Recently, blockchain has been introduced as an effective security solution deployed successfully in many applications in practice, such as, Bitcoin, cloud computing, and Internet-of-Things. However, the blockchain technology has not been adopted and implemented widely in m-commerce because its mining processes usually require to be performed on standard computing units, e.g., computers. Therefore, in this paper, we introduce a new m-commerce application using blockchain technology, namely, MobiChain, to secure transactions in the m-commerce. Especially, in the MobiChain application, the mining processes can be executed efficiently on mobile devices using our proposed Android core module. Through real experiments, we evaluate the performance of the proposed model and show that blockchain will be an efficient security solution for future m-commerce.

\end{abstract}



\section{Introduction}

The Technavio's market research analyst predicted that the global mobile commerce (m-commerce) market will grow at a compound annual growth rate of more than 27\% by 2020~\cite{Technavio}. M-commerce is gaining acceptance globally due to its variety of advantages, such as, offering multiple payment options to consumers, streamlining the procurement processes, and proving better price choices due to stiff market competition. However, the key concern about security for m-commerce transactions needs to be addressed to facilitate the continued growth of m-commerce.

Recently, blockchain has been introduced as an effective technology for data security problems. Furthermore, it has been implemented successfully in many applications, e.g., Bitcoin Wallet~\cite{ref_bitcoin_wallet}, Ethereum~\cite{ref_ethereum}, and Internet-of-Things (IoT)~\cite{ref_smarthouse}. Generally, blockchain is a distributed database that is replicated and shared among members of a network~\cite{ref_survey}. With blockchain, when a transaction is created, it will be verified parallelly and transparently by some nodes in the network through mining processes. After that, transactions are grouped into blocks, and the links between blocks and their content are protected by cryptography and cannot be forged. Once entered into a blockchain, transactions cannot be erased. Thus, a blockchain contains an accurate, time-stamped and verifiable record of every transaction, and hence the network does not need a central authority. As a result, the blockchain technology is popularly used in systems requiring high security and transparency, such as, Bitcoin~\cite{ref_bitcoin} and smart digital contract Ethereum~\cite{ref_ethereum}.

In blockchain technology, the mining process plays a crucial role in verifying and adding transaction records to the public ledger, i.e., the blockchain. In a mining process, the miner, i.e., the node taking the responsibility for mining a transaction, is required to verify the transaction and solve the proof-of-work problem in order to find a new hash for the new incoming block to store the verified transaction. This process is complicated and usually executed on powerful devices with high computational capacities and energy supply, e.g., servers and computers. Therefore, blockchain technology has not been adopted widely in m-commerce. However, the breakthrough of mobile technologies together with the improvement of hashing algorithms recently allow the mining processes to be able to implement on mobile devices efficiently, thereby opening new opportunities for the development of blockchain in m-commerce.

In this paper, we first introduce a new m-commerce application using blockchain technology, namely, MobiChain, to secure transactions. In MobiChain, we develop a Mobile Blockchain Application Programming Interface (API) which allows the mining processes to be performed on mobile devices effectively. We then conduct real experiments to evaluate the performance of the proposed module in terms of computation time, energy consumption, and memory utilization. Through experimental results, we show that blockchain technology will be an efficient security application for future m-commerce. To the best of our knowledge, our proposed Android core module is the first mobile application which allows performing mining processes on mobile devices. This application will be a pioneer tool enabling developers to implement blockchain applications on mobile devices in the near future. 

In Section~\ref{sec:RW}, we discuss related works. Then, the proposed system model and operations of the system are presented in Section~\ref{sec:SM&A}. After that we explain main functions of the core module in the MobiChain system. Section~\ref{sec:PE} describes how to deploy the system on real devices and shows important experimental results. Finally, conclusions and future works are presented in Section~\ref{sec:CL}.

\section{Related Work}
\label{sec:RW}

The success of Bitcoin has motivated researchers to investigate and adopt the blockchain technology to many areas, such as, smart contract~\cite{ref_ethereum}, finance~\cite{ref_bitcoin}, human resource management~\cite{ref_human}, supply chain~\cite{ref_supplychain}, and IoT~\cite{ref_smarthouse}\cite{ref_iot}. For example, in~\cite{ref_smarthouse}, the authors proposed a lightweight installation of the blockchain technology for a smart IoT house. In each house, multiple IoT devices (e.g., smart phones, personal desktops, and sensors) are connected to the same network. In addition, each house is equipped with an online, powerful resource device, which is referred to as a miner, taking responsibility to handle all transactions inside the house. Under the proposed model, only authorized users can access and control devices in the house, and messages received by the authorized users are protected and unable to be modified by any malicious users. The simulation results showed that the overheads of communications, processing, and energy consumption of the proposed approach are insignificant compared with the security and privacy gain by using blockchain. The authors in~\cite{ref_privacy} introduced a new decentralized personal data management platform that allows users to control their own data, thereby preventing the data breach from a third-party. The platform combines blockchain and off-blockchain storage to construct a personal data management framework. As such, the blockchain will recognize the users as the owners of their personal data, and thus the users are not required to trust any third-party. The authors in~\cite{ref_privacy2} then extended~\cite{ref_privacy} by improving the encryption algorithm and introducing a new method, called, Proof-of-Credibility Score (PCS), for mining processes. Different from~\cite{ref_privacy} where the trust score of the node is accumulated on how much good actions a node took, the proposed PCS method utilizes the connection between nodes to calculate the credibility score. The numerical results then showed that with the proposed hybrid blockchain of credibility score, the system security can be improved. 

As the blockchain application keeps growing, the scalability is a big challenge. The authors in~\cite{ref_bigchaindb} proposed a BigchainDB system using NoSQL database format to address this problem. The blockchain pipelining is adopted to allow the system to be scalable when adding blockchain-like characteristics to the distributed database. Experiment results showed that by using BigchainDB, the system can perform million writes per second, sub-second latency, and petabyte capacity. 

In m-commerce, transactions should be mined at mobile nodes to support direct peer-to-peer data exchange and sharing. This is especially important for the case when mobile nodes have no (or limited access to) Internet connection. Currently, there are some Android applications which was developed to perform mining processes on mobile devices for Bitcoin application, such as, Easy Miner~\cite{ref_easyminer} and LTC and Scrypt Miner PRO~\cite{ref_scr}. However, they are still demo versions and have not yet completed. More importantly, the platform to support general blockchain operations is still missing because Bitcoin applications are used for digital currency applications only. This is the motivation to introduce the MobiChain in this paper.

\section{System Model and Assumptions}
\label{sec:SM&A}

\begin{figure}[!t]
	\center\includegraphics[width=0.5\textwidth]{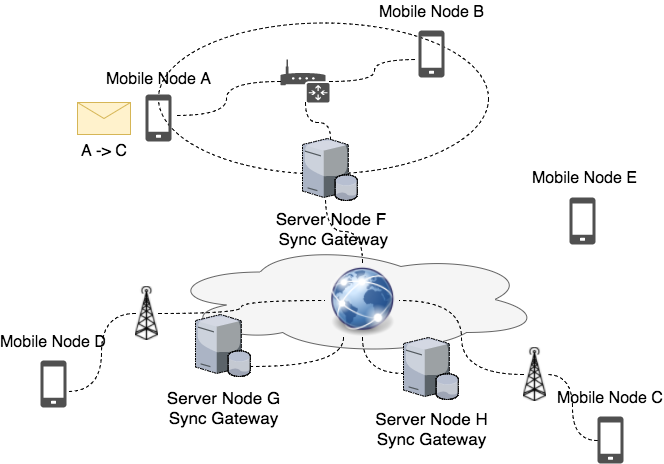} 
	\caption{The MobiChain system model.}
	\label{f_system}
\end{figure}

The proposed MobiChain system is illustrated in Fig.~\ref{f_system}, which consists of multiple mobile devices and multiple server nodes. Each mobile node has a backlog or queue to store pending tasks. The mobile node is connected to a server node either via the Internet or a local direct connection. All server nodes are installed \emph{Sync Gateway}~\cite{ref_sync_gateway} to broadcast data, and they are always connected to other server nodes via the Internet. In this case, we consider a message passing scenario, e.g., {\em Mobile~Node~A} sends a message to {\em Mobile~Node~C}. This message can be a data transfer request or a radio resource sharing request, which needs to be verified. This message is referred to as a transaction. We assume that all the mobile nodes are installed the MobiChain application with functions shown in Section~\ref{sec:Impl}, and they have already registered user accounts. These accounts are to provide public and private keys which are unique to the users. 

For notational convenience, let {\em M(X)} represent {\em Mobile~Node~X} ($ \forall X \in \{A,B,C,D,E\}$), and {\em S(Y)} represent {\em Server~Node~Y} ($\forall Y \in \{F,G,H\}$).

\subsection{Message Passing Process}

We consider a message passing process from {\em M(A)} to {\em M(C)}. This process includes the following steps:

\begin{itemize}
\item\textbf{Step 1: } The sender, i.e., {\em M(A)}, creates a transaction. In general, a transaction consists of three parts, i.e., text message, sender's information, and destination's information. After that, the sender puts the transaction into the sender's backlog.

\item\textbf{Step 2:} The sender uploads the transaction to its connected server, i.e., {\em S(F)}. 

\item\textbf{Step 3: } The Sync Gateway, i.e., {\em S(F)}, broadcasts the transaction to every node in the network. In Fig.~\ref{f_system}, {\em M(B)}, {\em M(C)}, and {\em M(D)} have the same transaction in their backlogs as they are connected to the network. Note that {\em M(E)} does not have the same transaction because it is currently not connected to the network. 
\end{itemize}

Here, we note that when node {\em M(E)} is reconnected to the network, it will be updated and synchronized with the current blockchain and transactions in the backlog. Currently, offline transactions are not considered, and this case will be studied in the future work.

\subsection{Mining Process}

After receiving the transaction from {\em M(A)}, we assume that nodes {\em M(B)} and {\em M(C)} are dedicated or they want to mine this transaction, e.g., they will receive some rewards to mine this transaction. These nodes will then start the mining process. The mining process includes five steps as illustrated in Fig.~\ref{f_mining}. The detail is as follows: 

\begin{figure}[!t]
\center\includegraphics[width=0.5\textwidth]{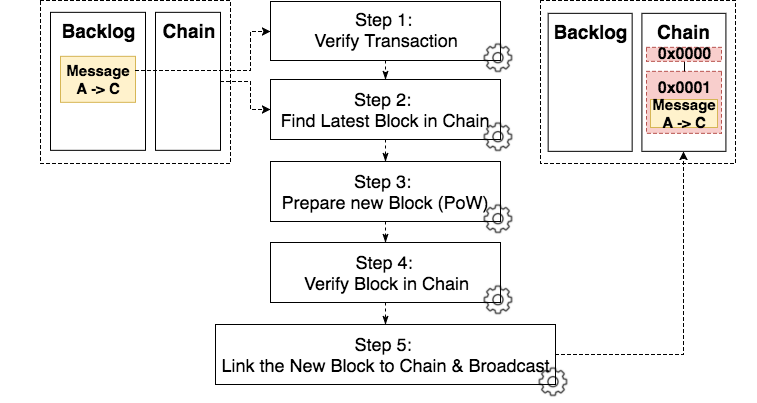} 
\caption{The mining process in the MobiChain system}
\label{f_mining}
\end{figure}
 
\begin{itemize}
\item\textbf{Step 1:} A miner, i.e., {\em M(B)} or {\em M(C)}, queries the transaction from its backlog, and verifies this transaction by checking whether the transaction is modified or not, and whether the transaction exists in the blockchain or not. If the transaction is not modified and does not exist in the blockchain, the miner performs Step 2. Otherwise, the mining process is terminated and the miner reports the problem to all other nodes in the network. 

\item\textbf{Step 2:} The miner finds the latest block in the chain, and stores the identification number (ID) of the latest block for the future use. The latest block may be referred to as the previous block. Note that if the chain is empty, the previous block ID is set to be zero. 

\item\textbf{Step 3:} 
This step is referred to as Proof-of-Work (PoW) process. In the PoW process, the miner will create a new block by hashing the information iteratively. The information includes (i) the IDs of the previous block and the created block, (ii) the block number (sequence number) of the created block, (iii) the verified transaction, (iv) the date/time stamp, and (v) the signature of the miner. We use \texttt{SHA3\_256} algorithm~\cite{ref_sha3} for the hashing process. 

\item\textbf{Step 4:} This step is to verify all information, e.g., hash ID and transaction information, in the chain. After creating the block ID, the miner verifies all the existing blocks in the chain. This process is to guarantee that the information of all blocks in the chain has not been changed.

\item\textbf{Step 5:} The miner needs to check whether the created block has been added to the chain by other mobile nodes or not. If not, the miner will add the created block to the chain. 
\end{itemize}

Note that in Fig.~\ref{f_mining}, the chain is empty. Hence, the identification of the previous block is zero, and the transaction that contains the message from {\em M(A)} to {\em M(C)} is pushed to a new block. The new block is then linked to the previous block in the chain.

%

\section{Implementation}
\label{sec:Impl}

The proposed core module is a Mobile Blockchain Application Programming Interface (API) for the Android environment. By using the proposed core module, all mining processes of the blockchain technology can be executed on Android devices. The functions of the core module are to create blocks, verify the correctness through mining processes, and link the verified blocks to the chain. Therefore, the core module requires an application to work as a front-end. In the previous section, the core module is implemented as a message passing application. In fact, the core module can be used in various applications, for example, file~sharing, smart contracts, and credit member systems. In this section, three functions of the core module are discussed in detail. 

\subsection{Database Function} 

The main functions of the chain, i.e., the database, are to record the private data in the local device and to broadcast this chain to all connected devices in the network. The chain includes three different data structures, i.e., account, transaction, and block structures, as presented in Fig.~\ref{f_accmodel}, Fig.~\ref{f_tranmodel}, and Fig.~\ref{f_blockmodel}, respectively. Note that the data structure is followed by the JavaScript Object Notation (JSON) format, and the explanation of each element is given in the comments. Moreover, a user has a set of public and private keys. The private key is stored locally at the mobile node, while the public key is broadcast to other mobile nodes as shown in the account data structure in Fig.~\ref{f_accmodel}. 

In the MobiChain system, the blockchain is stored in a database. The database is implemented on both mobile devices and servers. On the mobile devices, we use Coushbase Lite~\cite{ref_coucbbase_db} instead of the SQLite database available in the Android SDK environment because the Coushbase Lite is an embedded JSON database, which is a NoSQL system. NoSQL is a non-relational and schema-less data model, which makes NoSQL suitable for a real-time system. NoSQL also can handle partitioning of a database from multiple devices, and thus it is scalable. On the server nodes, Coushbase Sync Gateway~\cite{ref_sync_gateway} is implemented to receive and broadcast data to the devices. 
 
\begin{figure}[!t]
\centering
\begin{minipage}{24em}
\begin{lstlisting}[language=C]
{
	"type": "account",
	"username": /*String of username*/,
	"private_key": /*String of private key's account*/,
	"public_key": /*String of public key's account*/
	"create_date": /*Date time of creating*/
}
\end{lstlisting}
\end{minipage}
\vspace{-1em}
\caption{Account data structure followed by JSON format.}
\label{f_accmodel}
\centering
\vspace{-2em}
\begin{minipage}{24em}
\begin{lstlisting}[language=C]
{
	"id": /*Result string after hashing everything inside transaction excluding signature */,
	"signature": /*String of the combination between transaction and private key's sender*/,
	"timestamp": /*Time of creating*/,
	"transaction": {
		"data": {
			"payload": /*Any string in JSON format*/,
			"uuid": /*String of the unique identification number*/
		},
		"owner": [/*String of public key's sender*/, /*String of public key's destination*/]
	}
}
\end{lstlisting}
\end{minipage}
\vspace{-1em}
\caption{Transaction data structure followed by JSON format. }
\label{f_tranmodel}
\centering
\vspace{-2em}
\begin{minipage}{24em}
\begin{lstlisting}[language=C]
{
	"id": /*Result string after hashing block_number, tx_hash, previous_block, and nonce*/,
	"block_number": /*Integer of the current block number*/,
	"votes":[
				{
					"node_pubkey": /*String of public key's miner*/,
					"signature": /*Result string after vote is signed by using private key's miner*/,
					"vote": {
						"is_block_valid": /*Boolean that present the block valid status*/,
						"previous_block": /*String ID of the previous block*/,
						"timestamp": /*Time of block creating*/,
						"voting_for_block": /*Same with the ID*/
					}
				}
			],
	"version": "1",
	"tx_hash": /*Result string after hashing  all transactions in the block*/,
	"block": {
				"transactions": [/*list of transactions*/],
				"voters": [/*list public key's of voters*/]
	},
	"nonce": /*Integer of the hashing time. Note that hashing is done iteratively until the conditions are met*/
}
\end{lstlisting}
\end{minipage}
\vspace{-1em}
\caption{Block data structure followed by JSON format. }
\label{f_blockmodel}
\end{figure}

\subsection{Main Function} 

The Main Function is executed when one of the three following scenarios happens. 
\begin{itemize}
\item When a mobile node sends a new message, the Main Function creates a transaction and assigns the transaction to the sender's backlog. 

\item When a mobile node receives a transaction, the Main Function assigns the new transaction to the receiver's backlog. 

\item The Main Function is executed periodically to check whether a mobile node's backlog is empty or not. If the backlog is not empty, the main function will perform the mining process as discussed in Section III-B. 
\end{itemize}

In short, the Main Function will be used to create transactions and blocks, assign transactions into the backlog, and verify transactions in the backlog and blocks in the chain.

\subsection{Cryptography Function} 

The cryptography function can be separated into three parts, i.e., cryptography-hashes, key-signature, and encode-decode. In the cryptography-hashes, the SHA3-256 algorithm can be used~\cite{ref_sha3}. When a user registers for an account in the MobiChain system, a pair of private and public keys is generated by ED25519 public key signature system~\cite{ref_ed25519}, which will be used in signing and verifying a signature. For encode and decode processes, the Base58 schemes can be used~\cite{ref_base58}.

\section{Performance Evaluation}
\label{sec:PE}

All experiments presented in this section were performed on a Samsung Galaxy Tab S2 8.0 (T715), i.e., a mobile node, and a workstation with Intel Xeon CPU E5-1630, i.e., a Sync Gateway. The total energy consumption on the mobile device was measured by VideoOptimizer program~\cite{ref_videooptimizer}. The Mobilechain system and the core module were implemented by the Android Studio and Software Development Kits (SDK) tools. If the nonce is not specific in an experiment, the nonce is set to be zero. The nonce value is set in Fig.~\ref{f_blockmodel}.

In this section, we analyze memory utilization according to the number of blocks in the chain and the mining process performance. As shown in Fig.~\ref{f_mining}, the mining process has five steps. Most of the mining process computation is on Step~3 (chain verification) and Step~4 (Proof-of-Work), while those of Step~1, Step~2 and Step 5 are negligible. Therefore, only the chain verification and the Proof-of-Work processes are analyzed in this section. 

\subsection{Memory Utilization}

\begin{figure}[!t]
	\includegraphics[width=0.42\textwidth]{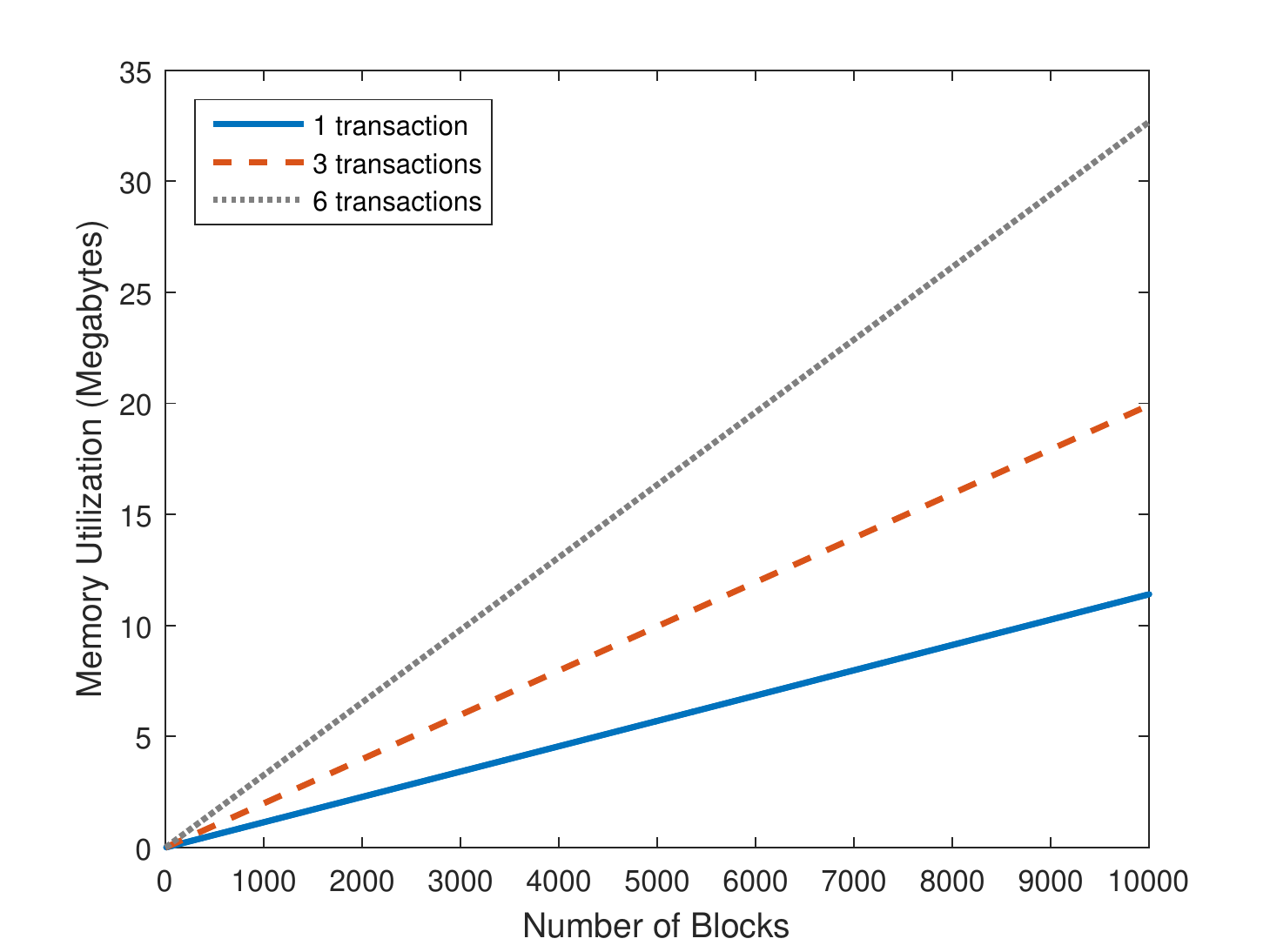}
	\caption{ The memory utilization when the number of blocks increases.} 
	\label{f_memory}
\end{figure}

Memory utilization of the proposed MobiChain is shown in Fig.~\ref{f_memory} under three different sizes of blocks, i.e., one transaction per block, three transactions per block, and six transactions per block. Here, the content of each transaction is fixed at 20 characters. In Fig.~\ref{f_memory}, if we increase the number of transactions in each block, the memory utilization can be reduced remarkably. In particular, if we store 3 or 6 transactions in one block, the memory utilization can be reduced by 33\% or 55\%, respectively. The reason can be explained by the following linear equation:
\begin{equation}
\text{Memory Utilization} = c_b + c_t T + c_d D 	,
\label{eq:MemoryUtilization}
\end{equation}
where $c_b$, $c_t$, and $c_d$ are constant, and they represent the size of block information, the size of one transaction, and the size of one digit, respectively. In~(\ref{eq:MemoryUtilization}), $T$ is the number of transactions in one block, and $D$ is the number of digits of $block\_number$ (as presented in Fig.~\ref{f_blockmodel}).

\subsection{Proof-of-Work Process}

In this experiment, the hash process is executed iteratively until the first three digits of the hash value equal zero. At each iteration, the hash value is combined with nonce, which will be increased by one in each iteration. The execution time of Proof-of-Work process is random due to the hash condition. The number of hash iterations until the condition is met is found to be random. 

\begin{figure}[!t]
	\includegraphics[width=0.42\textwidth]{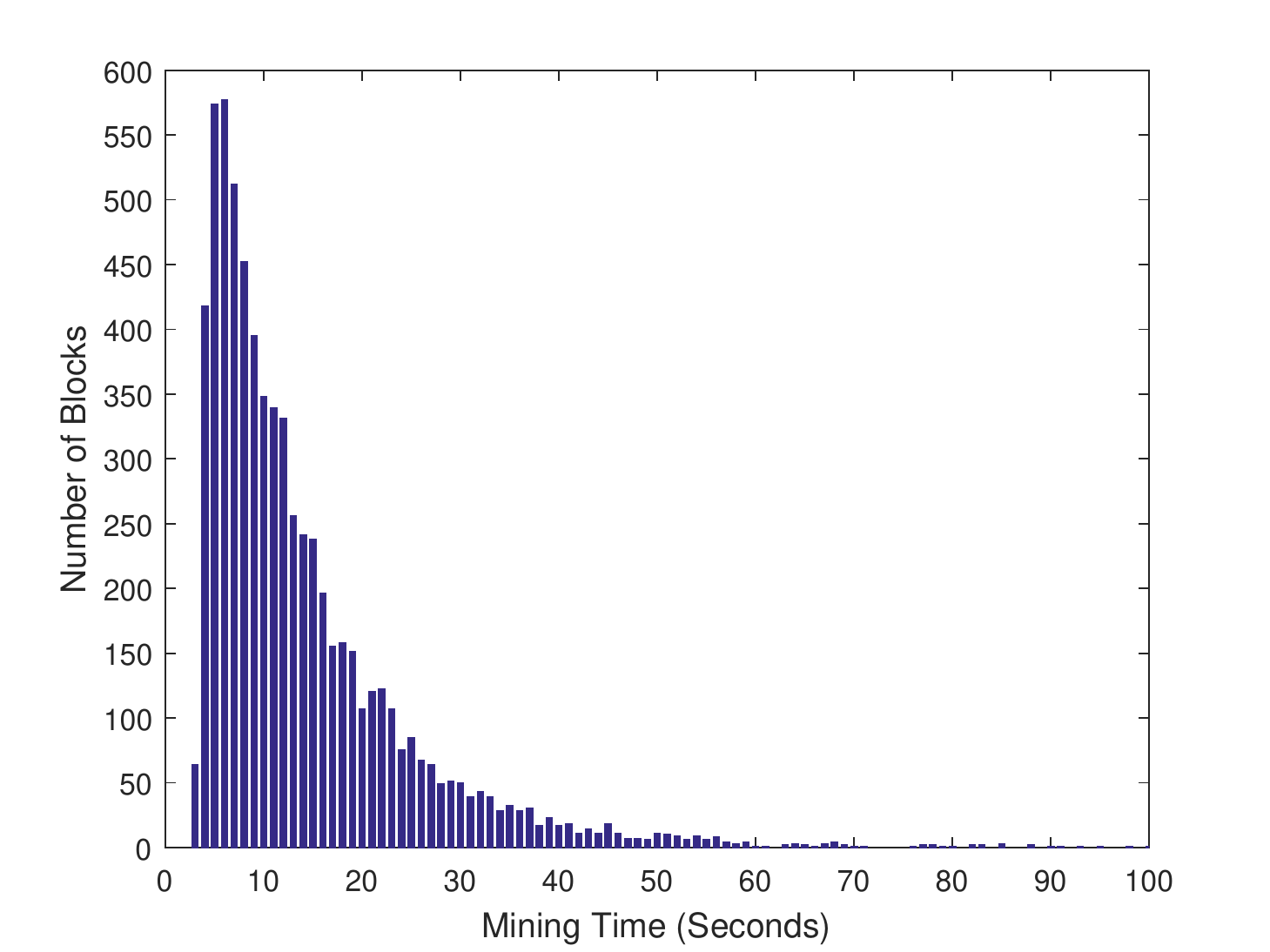} 
	\caption{The distribution of mining time v.s. number of blocks.}
	\label{f_hist}
\end{figure}

In our experiment, we create 7,156 blocks and use the mobile device to mine these blocks. In this case, it took 3.5 days to execute the Proof-of-Work processes for all 7,156 blocks. The histogram of these blocks is shown in Fig.~\ref{f_hist}, which can be expressed as a gamma long tail distribution~\cite{ref_gamma}. The experiment is filtered to show only 0 to 100 seconds. According to Fig.~\ref{f_hist}, $88.06\%$ of blocks need to use 3 to 30 seconds to perform the Proof-of-Work process, while only $4.79\%$ perform longer than 100 seconds. At the peak points, $23.23\%$ of the total blocks use 5 to 7 seconds. In our experiments, 803 hashing iterations are executed per second, and thus the peak points use around $4,015$ to $5,621$ hashing iterations before meeting the condition.

\subsection{Chain Verification Process}


\begin{figure}[!t]
\includegraphics[width=0.42\textwidth]{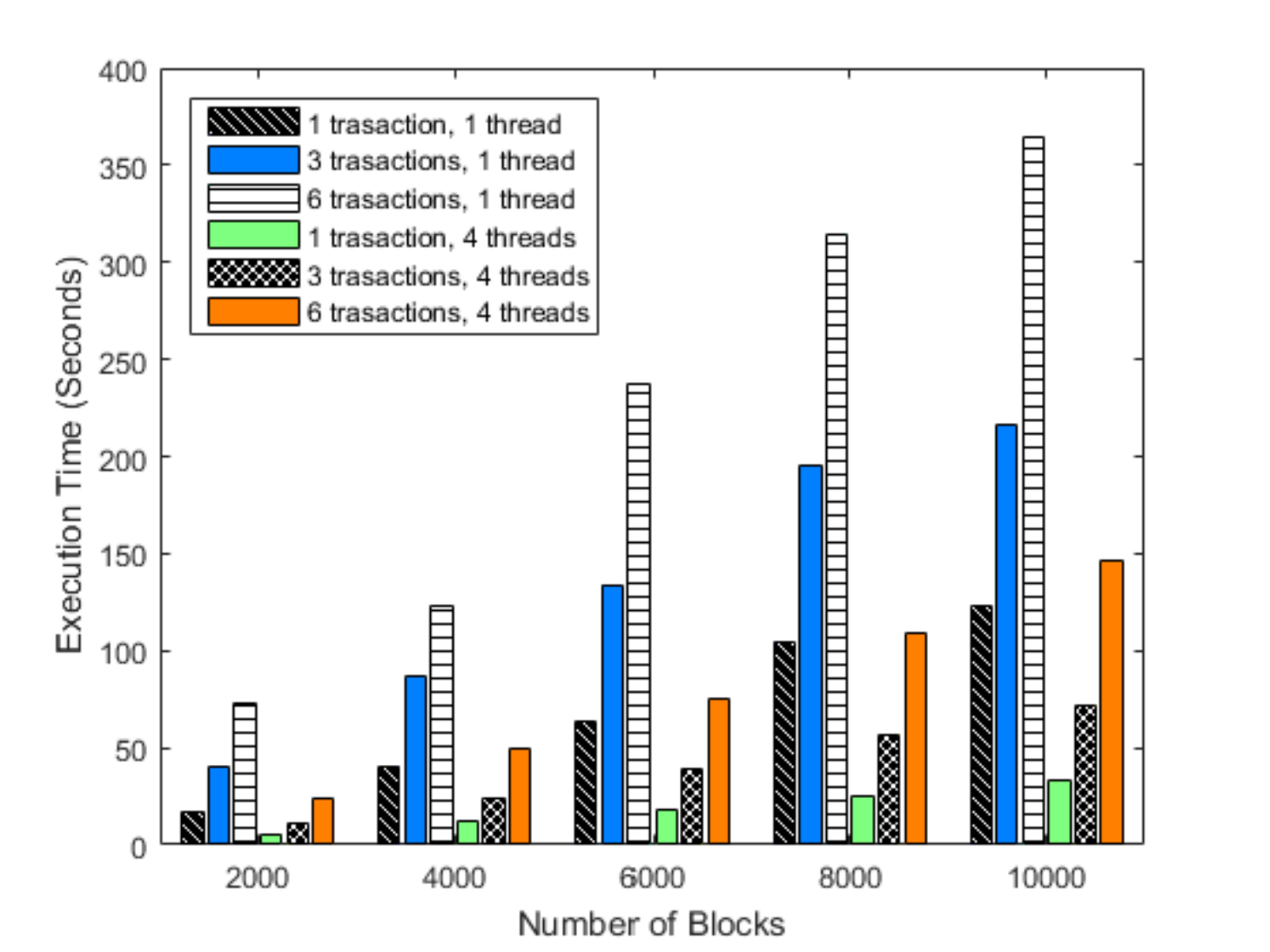} 
\caption{Execution time of the chain verification process. } 
\label{f_time}
\end{figure}

\begin{figure}[!t]
\includegraphics[width=0.42\textwidth]{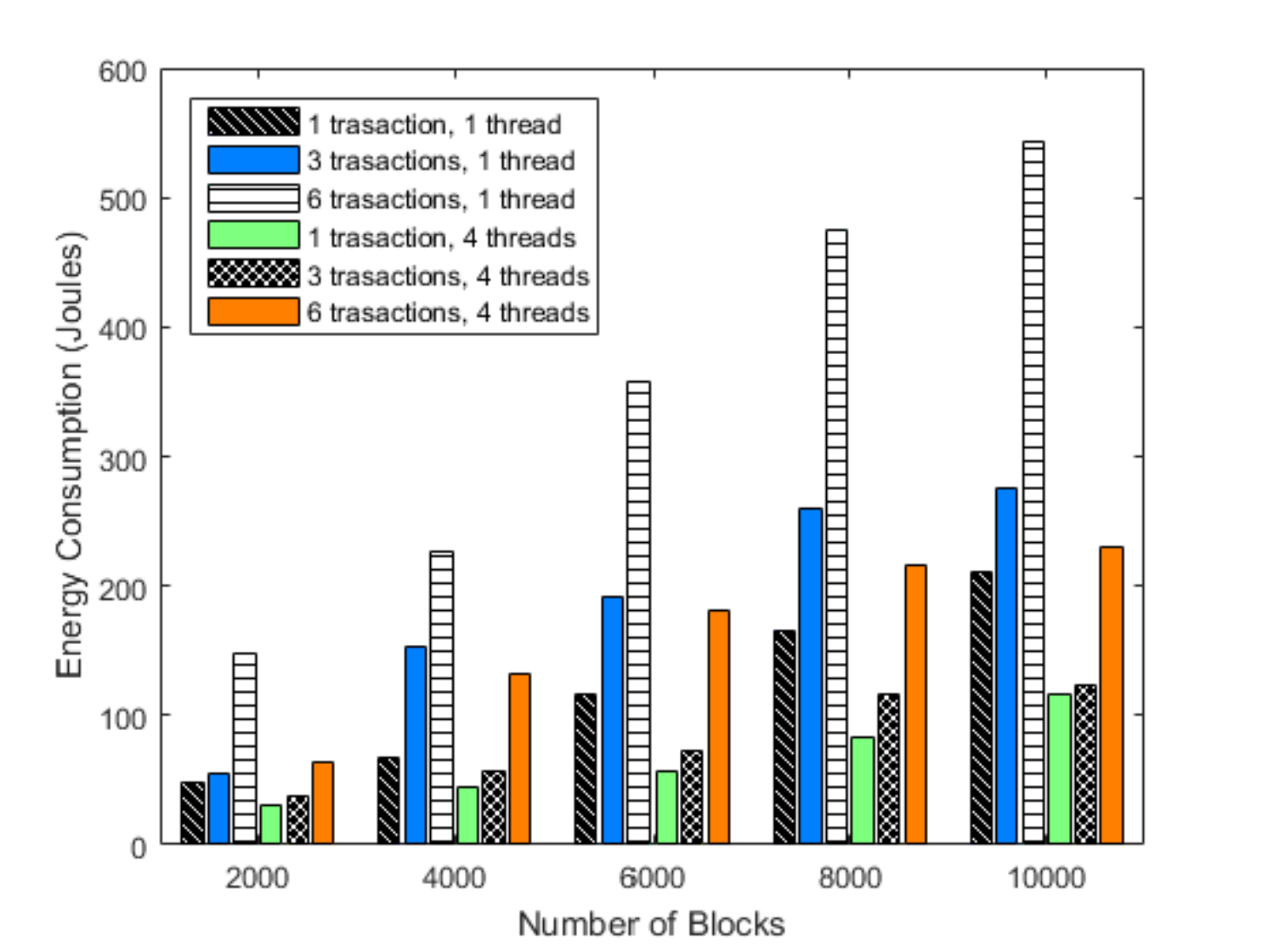} 
\caption{Energy consumption of the chain verification process}
\label{f_energy}
\end{figure}

\begin{figure}[!t]
\includegraphics[width=0.42\textwidth]{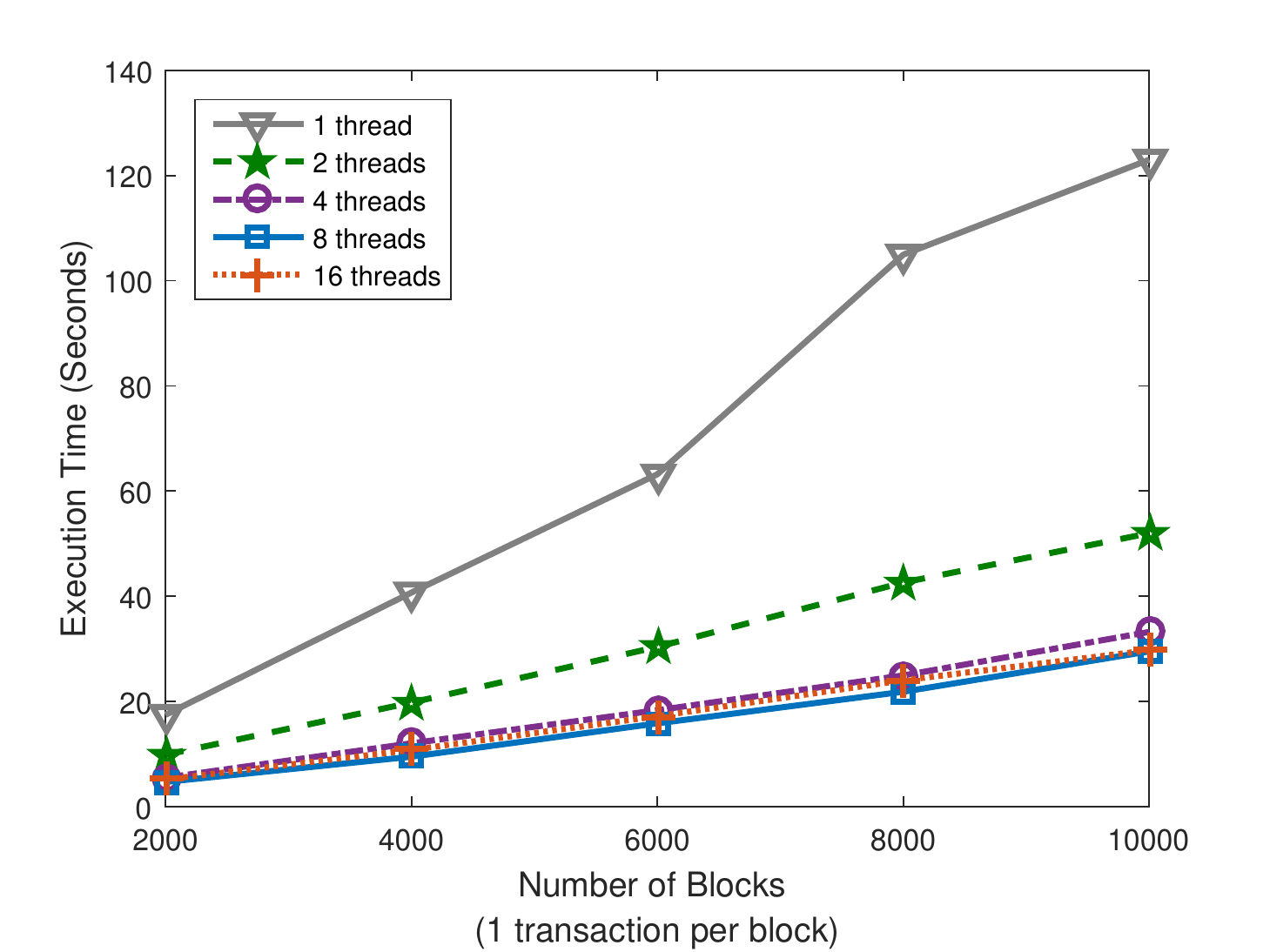} 
\caption{The execution time of the chain verification process by using multiple threads. } 
\label{f_multi}
\end{figure}

The execution time and energy consumption of the chain verification process are presented in Fig.~\ref{f_time} and Fig.~\ref{f_energy}, respectively. The execution time and energy consumption are measured from the beginning of the chain verification process until the end of this process. For multiple threads, the measurement is from the beginning until the last thread completes. Two types of experiments were conducted for both one-thread and four-thread scenarios. Each of them is performed with one, three, and six transactions per block. Each block contains 20 random characters. As expected, as the number of blocks in the chain increases, the execution time and energy consumption increase accordingly. 

One important finding is that the chain verification process is executed faster and consumes less energy when transactions are grouped together in a block. In particular, as shown in Fig.~\ref{f_time}, when there are 3 or 6 transactions grouped in one block, the total execution time per block can be reduced approximately 30\% or 40\%, respectively. When the execution time is reduced, the energy consumption will be decreased accordingly. However, in practice, having more number of transactions in a block can cause more delay if the transactions are generated randomly. Thus, the tradeoff for the number of transactions per block is worth investigating.

In Fig.~\ref{f_multi}, we show the execution time of one transaction per block with different threads. Clearly, as the number of blocks increases, the execution time rises. However, interestingly, when we increase the number of threads, the execution time is not always reduced. Specifically, when we increase the number of threads from 1 to 2 or from 2 to 4, the execution time is reduced approximately twice. However, if we keep increasing the number of threads, the execution time reduces insignificantly. Here, the best execution time is achieved when the number of threads is 8 because the Android device support 8 processing cores and each core has one thread.

\section{Conclusion}
\label{sec:CL}
In this paper, we have introduced MobiChain, a new m-commerce application using blockchain technology for data security. Different from most of current works, in our MobiChain system, mining processes can be performed on mobile devices through the proposed Android core module. This application provides useful functions for mobile transactions and promotes the development of m-commerce. Through experiment results, we have shown that blockchain technology is a practical solution for mobile devices to achieve security, efficiency, and scalability of data collection, processing, storage, and sharing. Furthermore, some important findings have been presented through experimental results, which are very useful for the development of blockchain technology in the future mobile applications. In the future, we will extend the MobiChain system for locally mining when the Internet is not available and propose data synchronization algorithms when mobile nodes are reconnected to the network, so that offline transactions can be performed accurately and efficiently.


\begin{thebibliography}{1}

\bibitem{ref_bitcoin_wallet} 
T. Bamert, C. Decker , R. Wattenhofer, and S. Welten, ``BlueWallet: The secure Bitcoin wallet,'' in \emph{International Workshop on Security and Trust Management}, pp. 65-80, Wroclaw, Poland, Sept. 2014.

\bibitem{ref_survey} 
I.-C. Lin and T.-C. Liao. ``A survey of blockchain security issues and challenges,'' \emph{International Journal of Network Security}, vol. 19, no.5, pp. 653-659, Sept. 2017.  

\bibitem{ref_smarthouse} 
A. Dorri, S. S. Kanhere, R. Jurdak, and P. Gauravaram, ``Blockchain for IoT security and privacy: The case study of a smart home,'' in \emph{IEEE International Conference on Pervasive Computing and Communications Workshops}, pp. 618-623, Hawaii, USA, Mar. 2017.

\bibitem{ref_privacy} 
G. Zyskind, O. Nathan, and A.S. Pentland, ``Decentralizing privacy: Using blockchain to protect personal data," in \emph{IEEE Security and Privacy Workshops}, pp. 180-184, San Jose, USA, May 2015.

\bibitem{ref_privacy2} 
D. Fu and L. Fang, ``Blockchain-based trusted computing in social network,'' in \emph{IEEE International Conference on Computer and Communications}, pp. 19-22, Chengdu, China, Oct. 2016. 

\bibitem{ref_human} 
X. Wang, L. Feng, H. Zhang, C. Lyu, L. Wang, and Y. You, ``Human resource information management model based on blockchain technology,'' in \emph{IEEE Symposium on Service-Oriented System Engineering}, pp. 168-173, San Francisco, USA, Apr. 2017.

\bibitem{ref_supplychain} 
H. M. Kim and M.Laskowski, ``Towards an ontology-driven blockchain design for supply chain provenance,'' Open-Access Online Library, Aug. 2016. Available at \url{http://dx.doi.org/10.2139/ssrn.2828369}

\bibitem{Technavio}
Global M-commerce Market 2016-2020. Available at: \url{https://www.technavio.com/report/global-media-and-entertainment-services-global-m-commerce-market-2016-2020}.

\bibitem{ref_bitcoin} 
S. Nakamoto, ``Bitcoin: A peer-to-peer electronic cash system'', White Paper, 2009. Available at: \url{https://bitcoin.org/bitcoin.pdf}


\bibitem{ref_iot} 
A. Dorri, S. S. Kanhere, and R. Jurdak, ``Towards an optimized blockChain for IoT,'' in \emph{IEEE/ACM Second International Conference on Internet-of-Things Design and Implementation}, pp. 173-178, Pittsburgh, USA, Apr. 2017. 

\bibitem{ref_ethereum}
V. Buterin, ``Ethereum: A next-generation cryptocurrency and decentralized application platform," Bitcoin Magazine, Jan. 2014. Available at: \url{https://bitcoinmagazine.com/articles/ethereum-next-generation-cryptocurrency-decentralized-application-platform-1390528211/}

\bibitem{ref_easyminer} 
Easy Miner, Available at: \url{https://play.google.com/store/apps/details?id=com.mr.app.ui&hl=en}

\bibitem{ref_scr}
LTC and Scrypt Miner PRO, Available at: \url{https://play.google.com/store/apps/details?id=com.miner.scrypt&hl=en}

\bibitem{ref_sha3} 
S.J. Chang, R. Perlner, W. E. Burr, M. S. Turan, J. M. Kelsey, S. Paul, and L. E. Bassham, ``Third-round report of the SHA-3 cryptographic hash algorithm competition," NIST Interagency Report 7896, Nov. 2012. Available at: \url{https://pdfs.semanticscholar.org/783c/a3d6f84102edd7597b50b92ed023ebbf669f.pdf}

\bibitem{ref_ed25519}
Ed25519: high-speed high-security signatures. Available at: \url{https://ed25519.cr.yp.to/}.

\bibitem{ref_base58} 
Base58Check encoding. Available at: \url{https://en.bitcoin.it/wiki/Base58Check_encoding}.

\bibitem{ref_coucbbase_db} 
D. Ostrovsky, M. Haji, and Y. Rodenski, ``Couchbase Lite on Android,'' in \emph{Pro Couchbase Server}, pp. 283-292, Publisher: Apress, Berkeley, CA, 2014.

\bibitem{ref_sync_gateway} 
D. Ostrovsky, M. Haji, and Y. Rodenski. ``Synchronizing data with the Couchbase Sync Gateway,'' in \emph{Pro Couchbase Server}, pp. 331-342, Publisher: Apress, Berkeley, CA, 2014.

\bibitem{ref_bigchaindb} 
T. McConaghy, R. Marques, A. Müller, D. De Jonghe, T. McConaghy, T, G. McMullen, and A. Granzotto, ``BigchainDB: a scalable blockchain database," White Paper, BigChainDB, 2016. 

\bibitem{ref_videooptimizer}
VideoOptimizer. Available at: {https://developer.att.com/video-optimizer/docs}

\bibitem{ref_gamma}
A continuous variable with a long tail distribution. Available at: \url{http://www.epixanalytics.com/modelassist/AtRisk/Model_Assist.htm#Extra_example_models/General/A_continuous_variable_with_a_long_tail_distribution.htm}


\end{thebibliography}
\end{document}